\documentstyle{mn}
\outer\def\gtae {$\buildrel {\lower3pt\hbox{$>$}} \over 
{\lower2pt\hbox{$\sim$}} $}
\outer\def\ltae {$\buildrel {\lower3pt\hbox{$<$}} \over 
{\lower2pt\hbox{$\sim$}} $}

\def\rchi{{${\chi}_{\nu}^{2}$}}

\title[Simultaneous observations of the near synchronous polar 
RX J2115--5840] {Simultaneous optical polarimetry and X-ray data of the near
synchronous polar RX J2115--5840}
\author[G. Ramsay et al]
{Gavin Ramsay$^{1}$, Stephen Potter$^{1,2}$, Mark Cropper$^{1}$, 
David A. H. Buckley$^{2}$, \and M. K. Harrop-Allin$^{1}$\\
$^{1}$Mullard Space Science Laboratory, University College London,
Holmbury St.Mary, Dorking, Surrey, RH5 6NT\\
$^{2}$South African Astronomical Observatory, PO Box 9, Observatory 7935, Cape
Town, South Africa\\
}

\date{Accepted MNRAS: 2 March 2000}

\begin{document}

\maketitle

\begin{abstract} 

We present simultaneous optical polarimetry and X-ray data of the near
synchronous polar RX J2115--5840. We model the polarisation data using
the Stokes imaging technique of Potter et al. We find that the data
are best modelled using a relatively high binary inclination and a
small angle between the magnetic and spin axes. We find that for all
spin-orbit beat phases, a significant proportion of the accretion flow
is directed onto the lower hemisphere of the white dwarf, producing
negative circular polarisation. Only for a small fraction of the beat
cycle is a proportion of the flow directed onto the upper hemisphere.
However, the accretion flow never occurs near the upper magnetic pole,
whatever the orientation of the magnetic poles. This indicates the
presence of a non-dipole field with the field strength at the upper
pole significantly higher. We find that the brightest parts of the
hard X-ray emitting region and the cyclotron region are closely
coincident.

\end{abstract}


\vspace{2.5cm}

\section{Introduction}

Polars (or AM Her systems) are amongst the most suitable objects with
which to study the interaction between an accretion flow and a
magnetic field. This is because the magnetic field of the accreting
white dwarf is strong enough ($B\sim$10--200MG) to prevent the
formation of an accretion disk. Therefore, the dominating emission
source at all wavelengths is the post-shock region above the surface
of the white dwarf. In polars, the spin period of the white dwarf and
the binary orbital period are generally synchronised and the accretion
flow from the dwarf secondary star threads onto magnetic fields lines
which have an unchanging orientation with respect to the white dwarf
(see Cropper 1990 and Beuermann \& Burwitz 1995 for general reviews
of polars).

However, four systems are known to be slightly ($\sim1\%$)
asynchronous (the near synchronous polars) and the accretion flow will
therefore attach onto different field lines as the flow rotates around
the white dwarf on the timescale of the spin-orbit beat period. Until
very recently, observations covering a beat period have been difficult
to obtain because the beat period is weeks or more (V1432 Aql: Watson
et al 1995, Friedrich et al 1996, Geckeler \& Staubert 1997; BY Cam:
Silber et al 1997, Mason et al 1998) or the system is faint (V1500
Cyg: Stockman, Schmidt \& Lamb 1988). Now a fourth system (RX
J2115--58) which is reasonably bright ($V\sim$17) has been discovered
with a beat period of 6.3 days (Schwope et al 1997, Ramsay et al 1999)
which allows a detailed study of these systems to be undertaken for
the first time.

The observations of Ramsay et al (1999) made over 13 nights in 1997
provide the most direct evidence that the accretion flow is directed
onto one magnetic pole and then the other as the flow rotates around
the white dwarf on the timescale of the beat period. This is most
apparent in the circular polarisation data which shows a change of
sign when the accreting flow is directed onto opposite magnetic
poles. From these observations they determined the spin period of the
white dwarf to be $P_\omega$=109.55 mins and the binary orbital period
to be $P_\Omega$=110.89 mins.

The observations of Ramsay et al (1999) were difficult to reconcile
with simple views of how the accretion stream attaches onto the
magnetic field of the white dwarf. To investigate this in greater
detail we have obtained simultaneous optical polarisation and X-ray
data obtained using {\sl RXTE} in July 1998.

\section{Observations}

\subsection{Polarimetry Data}

White light optical polarimetry data were obtained using the SAAO 1.9m
telescope at Sutherland, South Africa and the UCT Polarimeter (Cropper
1985). Table \ref{log} details these observations. There were
occasions when thin cloud was present which compromised some of the
photometry but did not effect the polarisation data. The data were
reduced as described in Cropper (1997).

\begin{table}
\begin{tabular}{lrrr}
\hline
Date&Telescope&HJD  &Duration\\
    &         &Start& \\
\hline
Optical Polarimetry& & & \\
\hline
1998 July 21& SAAO 1.9m&16.37& 7h 20m\\
1998 July 22& SAAO 1.9m&17.36& 6h 25m\\
1998 July 23& SAAO 1.9m&18.34& 5h 30m\\
1998 July 24& SAAO 1.9m&19.31& 7h 25m\\
1998 July 25& SAAO 1.9m&20.32& 8h 20m\\
1998 July 26& SAAO 1.9m&21.31& 8h 35m\\
\hline
X-ray& & & \\
\hline
1998 July 21&{\sl RXTE}& 16.59& 7.6ksec\\
1998 July 22&{\sl RXTE}& 17.59& 7.9ksec\\
1998 July 23&{\sl RXTE}& 18.53& 13.2ksec\\
1998 July 24&{\sl RXTE}& 19.55& 6.9ksec\\
1998 July 25&{\sl RXTE}& 20.53& 13.1ksec\\
1998 July 26&{\sl RXTE}& 21.57& 6.9ksec\\
1998 July 27&{\sl RXTE}& 22.50& 14.2ksec\\
\hline
\end{tabular}
\caption{Observing log for RX J2115--580. The HJD the start time is
HJD+2451000.0.}
\label{log}
\end{table}

Ramsay et al (1999) made the point that the circular polarisation data
is the most useful aspect of the polarisation data because the sign
change of the circular polarisation data indicates a change in
accreting pole. Taking this lead we initially concentrate on the
circular polarisation data.

\begin{figure}
\begin{center}
\setlength{\unitlength}{1cm}
\begin{picture}(8,12.8)
\put(-1.5,-2.0){\includegraphics{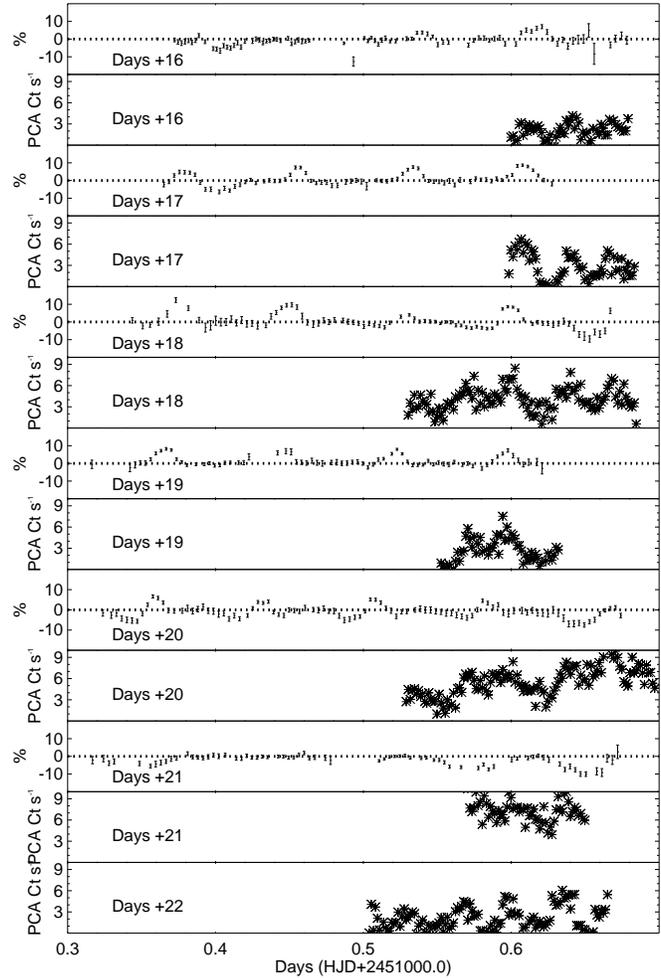}}
\end{picture}
\end{center}
\caption{The circular polarisation data and {\sl RXTE} PCA (2--15keV)
data. The HJD is shown on the left hand side +2451000.0.}
\label{cpol1} 
\end{figure}

As in the 1997 data, our new data have some nights which show positive
and negative polarisation and also other nights when only one sign of
polarisation was seen (Fig. \ref{cpol1}). However, in contrast to the
1997 data, which were reasonably repeatable over the course of one
night, the new data are more variable. For instance, the peak in
negative circular polarisation seen at HJD=2451016.4 is not repeated
during the night. More interestingly, a negative circular polarisation
peak is expected at HJD=2451021.42 based on the timing of the peaks
that night; however, none is seen. Checks (such as determining how the
sky polarisation varied over the course of the night) were made to
determine if this could be due to instrumental problems, but we have
ruled these out.  The lack of a polarisation peak therefore implies
that the accretion flow is being directed equally towards both poles
and the net polarisation is zero, or that the accretion flow is very
unstable and can `switch-off' on timescales as short as one orbital
period. The fact that there is no dip in the intensity curve at this
point suggests that the former scenario is more likely.

\begin{figure*}
\begin{center}
\setlength{\unitlength}{1cm}
\begin{picture}(12,15.5)
\put(-2,-10.5){\includegraphics{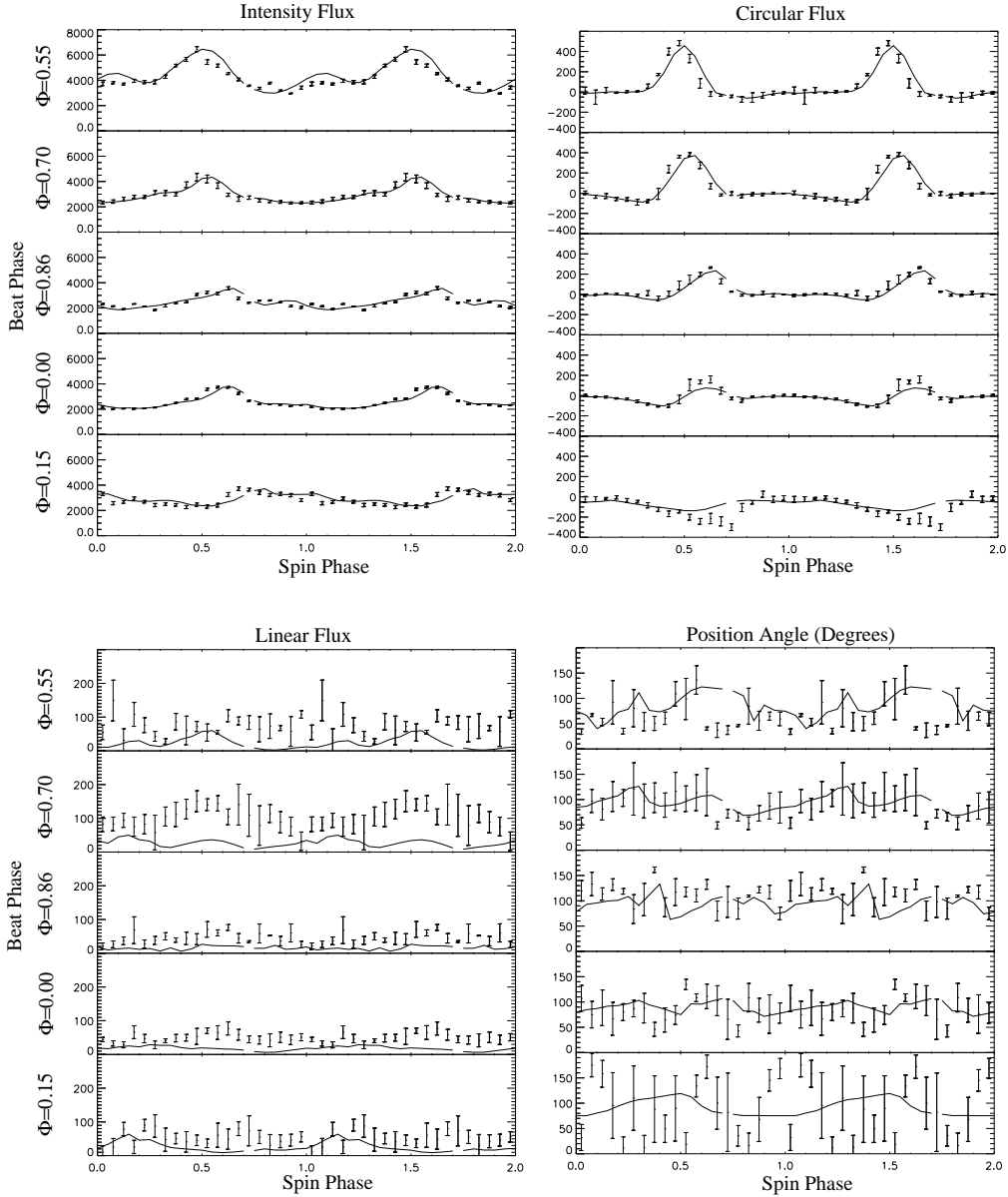}}
\end{picture}
\end{center}
\caption{The polarisation and intensity data folded on the spin period
together with the model fits for various spin-orbit beat phases. Phase
zero of the beat phase has been chosen so that it matches
approximately that of beat phase of Ramsay et al (1999). 
The phasing of the spin period is arbitrary.}
\label{pol_model} 
\end{figure*}

A Discrete Fourier Transform (DFT) was used to search for periodic
signals in the data. The amplitude spectrum is similar to that of the
1997 data (Fig. 2 of Ramsay et al 1999) although the resolution is
lower since only 6 nights are covered.  The amplitude spectrum was
pre-whitened using the spin and orbital frequencies (and their first
harmonics and $P_{\omega} \pm P_{\Omega}$ side bands) detected in the
1997 data and shown in Table 1 of Ramsay et al (1999). There were no
remaining frequencies with an amplitude above 0.1 percent and we find
no evidence for any other side band frequencies. To examine how the
polarisation data varies over the spin-orbit beat period, we show in
Figure \ref{pol_model} these data folded on the spin period as a
function of the beat period. Since we do not know the beat period
accurately enough to phase our new data with that of our 1997 data, we
have matched the beat phase of our new data so that it approximately
matches that of the 1997 data (to within $\sim$0.1 cycles). The
phasing of the spin phase is arbitrary. We discuss these data in
detail in \S \ref{polmod}.

\subsection{X-ray data}

RX J2115--58 was detected in the 2-15 keV range using {\sl RXTE} when
it was observed in 1998 July 21--27 (Table \ref{log}). To improve the
signal-to-noise ratio, we extracted data from only the top Xenon layer
of the PCA. The background was estimated using {\tt PCABACKEST} V2.1b
using the faint source models applicable to the date of the
observation. After excluding data contaminated by high background
rates, we were left with 65 ksec of data with a mean background
subtracted count rate of 3.9 ct s$^{-1}$.

The background subtracted light curve using the full 2-15 keV range is
also shown in Fig. \ref{cpol1}. Unlike many polars the amplitude of
the variation is comparatively small. There are, however, bright peaks
in the light curve which correspond to peaks in the white light
intensity curve and the peaks in the circular polarisation curve.

As for the circular polarisation data we used a DFT to search for
periodic modulations in the X-ray light curve.  A peak is found which
corresponds to the spin frequency. Its amplitude is much less
prominent compared to the amplitude spectrum of the circular
polarisation data. The data were pre-whitened using the spin and
orbital frequencies (and their first harmonics and $P_{\omega} \pm
P_{\Omega}$ side bands) as for the optical data: no additional
significant peaks were found.

\begin{figure*}
\begin{center}
\setlength{\unitlength}{1cm}
\begin{picture}(8,10)
\put(-5.5,-28){\includegraphics{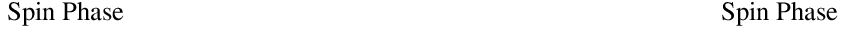}}
\end{picture}
\end{center}
\caption{The X-ray data folded on the spin period as a function of 
the spin-orbit beat period for two energy bands. These data have been
phased in the same manner as our optical data.}
\label{xbeatfold} 
\end{figure*}

To examine the intensity variation as a function of the beat period,
we show in Fig \ref{xbeatfold} the X-ray data folded on the spin
period as a function of the beat period. The shape of the spin-folded
light curve varies quite considerably during the beat cycle. The
modulation is rather low at beat phase $\psi$=0.35 and 0.72. On the
other hand a large peak is seen in the spin-folded data at beat phase
$\psi$=0.56. We searched for any hardness variation in the spin-folded
light curves by dividing the 2--4keV folded and binned data with the
equivalent 4--15keV data for all beat phases: no evidence was found
for a significant variation. We go onto discuss the X-ray spectrum of
RX J2115--58 in more detail in \S \ref{xfit} and possible explanations
for the change in the spin-folded light curves in \S \ref{shift}.

\section{The spin and orbital periods of RX J2115--58}

Determining accurate orbital and spin periods of near synchronous polars
is far from trivial. This is because if the accretion flow is directed
onto one then the other pole during the spin-orbital beat cycle -- a
`pole-switch' -- different periods can be obtained for observations
covering different fractions of the beat cycle (Mouchet et al 1997).

The most comprehensive set of observations of RX J2115--58 is that of
the polarimetry data of Ramsay et al (1999) which covered a time
interval of 13 days. They determined a spin period, $P_\omega$=109.55
mins and an orbital period of $P_\Omega$=110.89 mins.  This orbital
period is very similar to that determined by Vennes et al (1996)
(110.8 mins) who obtained 4 nights of spectroscopy over a time span of
6 days. Another slightly longer orbital period (111.3 mins) was
determined by Buxton et al (1999) who obtained spectroscopic data on 5
consecutive nights. To further complicate matters, Schlegel (1999)
analysed {\sl ROSAT} HRI data and tentatively found 2 periods, 102.6
and 110.4 mins which did not coincide with any other period
determinations.

We do not place great weight to the periods determined by Schlegel
(1999) since they were determined from data lasting only 14.8 ksec
(just over 2 spin cycles) which was spread over 59 ksec. On the other
hand, the period determined by Buxton et al (1999) deserves further
comment. In an analysis of the near synchronous polar BY Cam (with a
beat period of 14.5 days), Mouchet et al (1997) found that the period
determinations in that system were biased depending on the time
interval over which they were collected. It is likely that this can
account for the difference between the orbital period determinations
of Buxton et al (1999), Vennes et al (1996) and Ramsay et al
(1999). Since the polarisation data of Ramsay et al (1999) covers the
longest time interval ($\sim$1.4 beat cycles), it is probable that
these period determinations have the least bias.

Although the amplitude spectra of the X-ray and polarimetric data
presented in this paper are consistent with the periods determined by
Ramsay et al (1999), we must regard these periods with some degree of
caution until they can be confirmed. Nevertheless, to make progress,
we assume these periods for the rest of this paper. Unfortunately,
these periods are not precise enough to phase together the data in
Ramsay et al (1999) and the data presented here.

\section{Modelling the Polarimetry data}
\label{polmod}

\subsection{Background}

To model our polarisation data we used the optimisation method (Stokes
imaging) of Potter, Hakala \& Cropper (1998). This method finds the
best fit to the data and maps the shape, location and structure of the
cyclotron emission region(s) in an objective manner. Stokes imaging
finds the smoothest solution using a regularisation technique. It
has been successfully used to fit data on other polars, eg MN Hya
(Ramsay \& Wheatley 1998), V347 Pav (Potter et al 2000) and ST LMi
(Potter 2000).

Strictly speaking, the model maps the accretion region(s) in terms of
an optical depth parameter. In this paper we assume a simple dipole
field and the cyclotron model at the core of the Stokes imaging is a
10 keV constant temperature model of Wickramasinghe \& Meggitt (1985)
which is suitable for modelling white light observations (Potter
2000). We used a dipole magnetic field strength of 15 MG. This is
consistent with that found by Vennes et al 1996 (\ltae 20 MG) and
Schwope et al 1997 (11$\pm$2 MG). In fact, the resulting fits to white
light data are not affected by small changes to the magnetic field
strength (Potter 2000).

The aim of the Stokes Imaging technique is to produce highly resolved
images of the emission regions as they change from one beat phase to
the next. However, in the case of our data, the time resolution of the
data is only 20 points per spin phase of the white dwarf. Therefore
one data point corresponds to 18 degrees of rotation by the white
dwarf. Thus the maximum resolution of the map would be 18 square
degrees. In practice, it is slightly worse than this because of the
regularisation term which smooths the images, and also because the
amount of polarised flux, especially the linear flux, is not
particularly high. This will therefore result in blurred images of the
cyclotron emission regions. However, the main aim here is to
investigate how the emission regions move across the surface of the
white dwarf from one beat phase to the next. Thus, even though the
image resolution is quite low, any systematic movement of the region
across the surface of the white dwarf will be evident in the cyclotron
emission maps.

\subsection{Preliminary fits to the data}
\label{prelim}

As the system is near synchronous we need to construct a different
accretion map for each beat phase. Figure \ref{cpol1} shows that even
during the course of a single night the accretion geometry can
change. To obtain accurate maps of the accretion region(s) we need
good phase resolution and small errors on the folded and binned data
points.  We therefore have chosen time intervals in which the
polarisation data was repeatable. Naturally, in those intervals when
the data was not repeatable, this would lead to significant changes in
the resulting accretion maps. For instance at HJD=2451021.42 (\S 2.1
and Figure 1) where a negative circular polarisation peak was not seen
when expected, the accretion map would be very different at that
epoch. We return to this point in \S \ref{shift}.

Since this system is not eclipsing and there has been no detailed
modelling of polarisation data until now to give us an indication of
the system geometry, we initially had to search a wide range of
parameter space. The binary inclination, $i$, was searched between
$i=$0--80$^{\circ}$, the angle between the binary inclination and the
magnetic axis, $\beta$, was searched between $\beta$=0--90$^{\circ}$, and
the phase at which magnetic pole crosses our line of sight, $\zeta$,
was searched between $\zeta$=0--360$^{\circ}$.

For all the beat phase intervals that we observed, we found the best
fit solutions had high inclinations, low dipole offsets and the phase
at which the magnetic pole crosses our line of sight $\zeta$
corresponds to spin phase 0.25. Outside this parameter range the fits
were very poor.  Although $i$ and $\beta$ are not strongly constrained
we can rule out low inclinations and high dipole offsets.  The data
are not of sufficient quality to allow us to constrain the angles
further.

In our fitting routine all four Stokes parameters are equally
weighted. However, because the sign of the circular polarisation is an
indicator of which magnetic hemisphere of the white dwarf accretion is
occurring, we consider the fit to the circular polarisation data to be
the most important parameter when it comes to examining the fits by
eye. In addition, unlike the intensity data (which will be
contaminated by other non-accretion driven sources) it is pure
cyclotron emission. Upon inspection of the final fitness values of all
the solutions (a parameter which indicates the goodness of fit and the
smoothness of the map, Potter, Hakala \& Cropper 1998), we find that
the fits to the circular polarisation are significantly better for
inclinations above 60$^{\circ}$ and dipole offsets less than
20$^{\circ}$.  In the next stage of our modelling we kept these
parameters fixed at the following values for each beat phase:
$i=70^{\circ}$, $\beta=10^{\circ}$, $\zeta= 90^{\circ}$. These values
are not necessarily the best values for a particular beat phase, but
they are however the best values that give good fits for all beat
phases.

\subsection{Detailed fits to the data}

\begin{figure}
\begin{center}
\setlength{\unitlength}{1cm}
\begin{picture}(6,18)
\put(-6.5,-7){\includegraphics{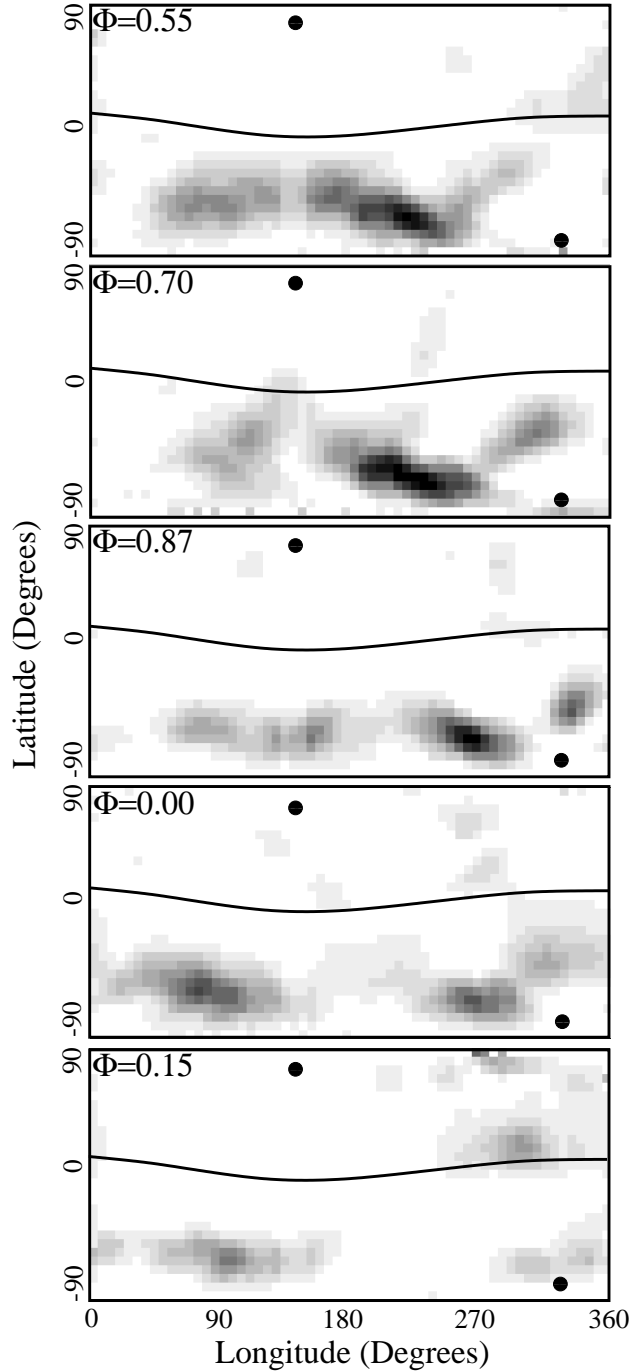}}
\end{picture}
\end{center}
\caption{The accretion regions on the white dwarf are shown in spin
coordinates as a function of beat phase. The dark dots mark the
locations of the magnetic poles and the magnetic equator is shown as a
solid line. The upper magnetic pole points most favourably
to the observer at spin phase $\phi$=0.25 (\S \ref{prelim}).}
\label{pol_model2} 
\end{figure}

\begin{figure*}
\begin{center}
\setlength{\unitlength}{1cm}
\begin{picture}(12,6)
\put(-5,11.){\includegraphics{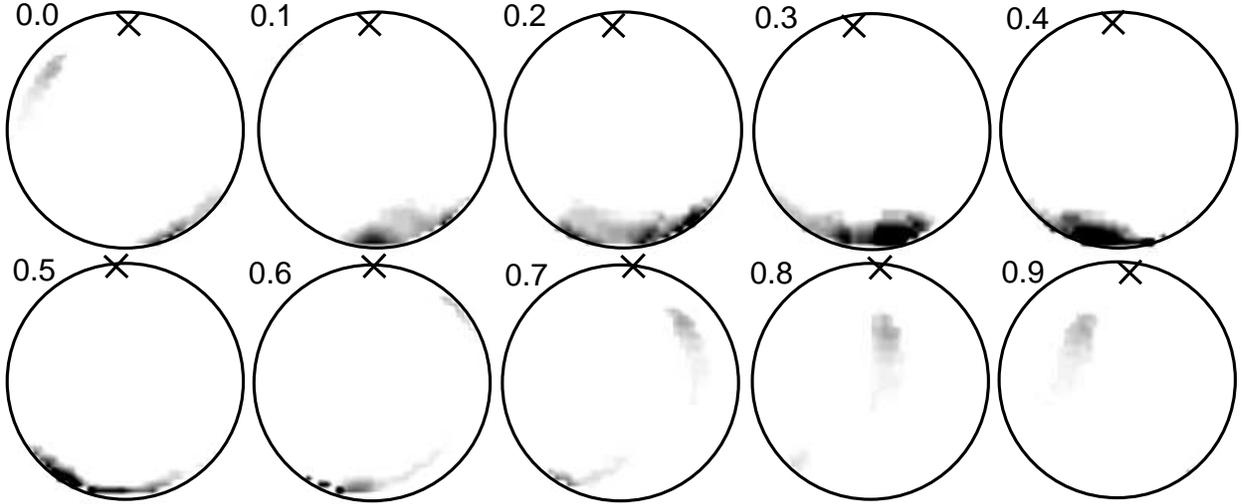}}
\end{picture}
\end{center}
\caption{The location of the accretion regions for the spin-orbit beat
phase  $\psi$=0.55 as predicted by our model. The spin phase of the
white dwarf is shown while the 
cross marks the location of the upper magnetic pole.}
\label{globes1} 
\end{figure*}

\begin{figure*}
\begin{center}
\setlength{\unitlength}{1cm}
\begin{picture}(12,6)
\put(-5,11.){\includegraphics{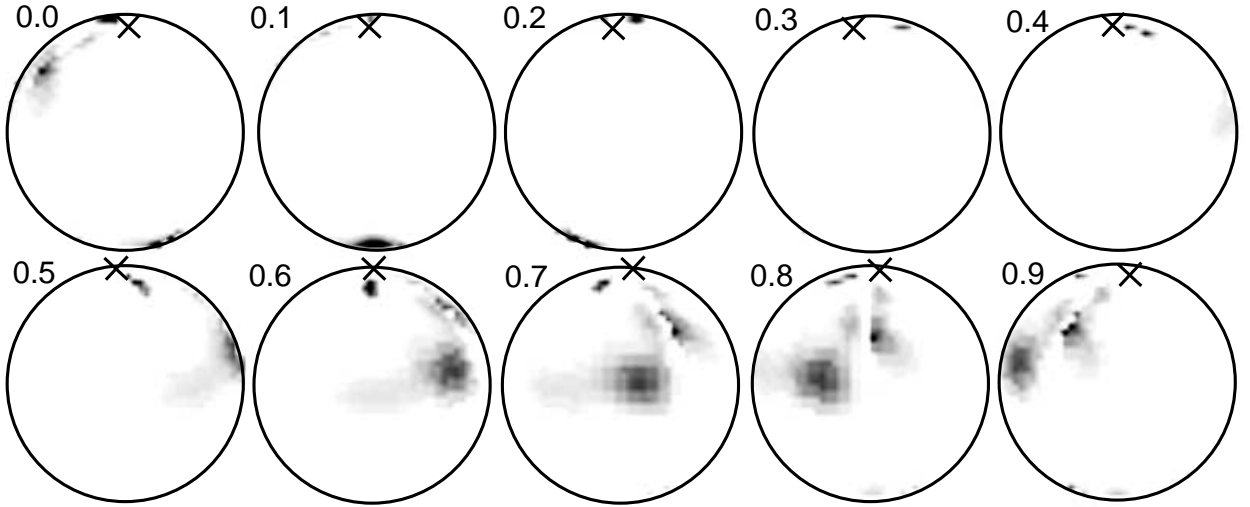}}
\end{picture}
\end{center}
\caption{As Figure \ref{globes1} but for the spin-orbit beat
phase  $\psi$=0.15.}
\label{globes2} 
\end{figure*}

The model polarisation curves along with the data are shown in
Fig. \ref{pol_model}. The fits to the intensity data are good, as are
the fits to the circular polarisation with the exception of the beat
phase $\psi$=0.15.  On the other hand the linear polarisation data is
consistently underestimated in our model. This is a result of two
effects. Firstly the lower count rate in the linear flux and hence the
relatively larger error bars compared to the intensity and circular
flux will result in the linear flux not having a large effect in the
fitting procedure. Secondly, the large size of the emission
regions that have arisen due to the low time resolution of the data
will lead to the smoothing of any linearly polarised features.  The
emission maps are shown in a Mercator projection in
Fig. \ref{pol_model2} for 5 spin-orbit beat phases. We also show the
emission regions projected onto a globe as a function of spin phase
for two spin-orbit beat phases in Fig. \ref{globes1} \& \ref{globes2}.

The maps shown in Fig. \ref{pol_model2} show how the location of the
emission regions change over the beat cycle. The location of the
magnetic poles and the equator are also shown. Emission regions on
opposite magnetic hemispheres will, of course, exhibit circular
polarisation of opposite sign when viewed at the same orientation. For
the majority of the beat period which we were able to observe,
significant emission is present only in the lower hemisphere. Only at
beat phase $\psi$=0.15 is there significant emission present in the
upper hemisphere -- this is the only beat phase where there is a
negative peak in circular polarisation (Fig. \ref{pol_model}). All
other beat phases have positive circular polarisation peaks. We
therefore conclude that cyclotron emission from the lower hemisphere
is positively circularly polarised while in the upper hemisphere it is
negatively polarised.

Because of the orientation of the binary system, the accretion region
in the lower hemisphere appears at the edge of the visible disk of the
white dwarf (Fig. \ref{globes1} \& \ref{globes2}). At beat phase
$\psi$=0.15 this region is visible for only a very short proportion of
the spin cycle and the majority of the accretion is directed onto the
upper hemisphere. Caution should be exercised when interpreting the
structure of the accretion region in the upper hemisphere
(Fig. \ref{globes2}) since the fit to the circular polarisation data
is rather poor between spin phases $\phi\sim$0.5--0.8 when this
accretion region is in view. All we can say with confidence is that
accretion is taking place onto the upper hemisphere at these spin
phases rather than the lower hemisphere. We discuss the results of
these polarisation maps, together with the X-ray light curves, in more
detail in \S \ref{shift}. However, we now go onto discuss the X-ray
light curves with these accretion maps in mind.

\section{Modelling the X-ray data}
\label{xfit}

It is reasonable to assume that the polarised optical flux and the
hard X-ray flux originate from the same shock region above the white
dwarf. It should be possible, therefore, to use the cyclotron opacity
maps derived in the previous section as an indication of the hard
X-ray emission region. We have therefore constructed a simple model
that uses the cyclotron maps to produce the hard X-ray emission one
would expect from such an region, assuming optically thin X-ray
emission. This is only possible because we have quasi-simultaneous
optical and X-ray data. However, for beat phases $\psi$=0.35 and
$\psi$=0.40 this is not possible since we do not have the cyclotron
opacity maps for these beat phases. We also assume that the map of the
optical depth parameter is an indication of the density of the X-ray
emitting region and, hence, higher densities give more hard X-ray flux.

\begin{figure}
\begin{center}
\setlength{\unitlength}{1cm}
\begin{picture}(8,10)
\put(-3,-3.8){\includegraphics{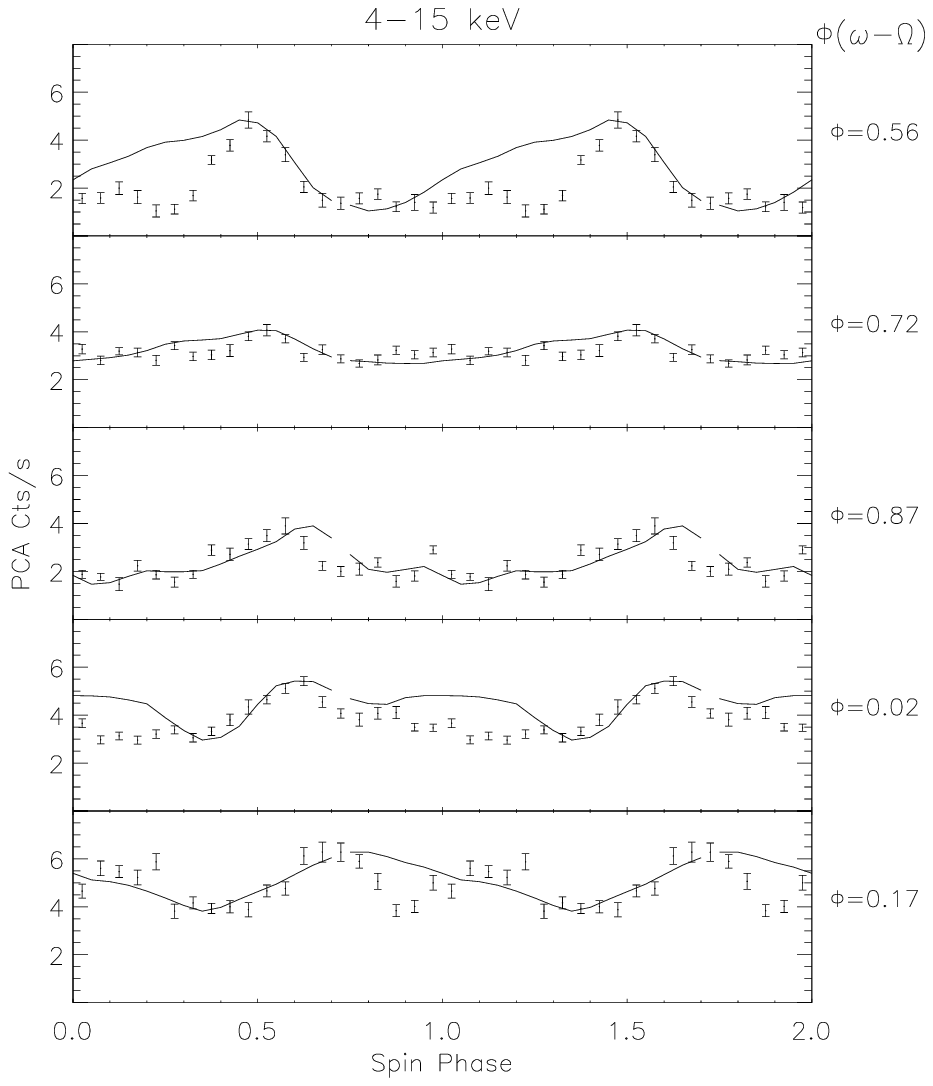}}
\end{picture}
\end{center}
\caption{The X-ray data (4--15keV) overlaid with the model X-ray light
curve obtained using the maps of the location and extent of the
emission regions derived using the polarisation mapping. The close
agreement between the phasing of the peaks of the data and model
suggest a close coincidence between the brightest part of the hard
X-ray and cyclotron emitting regions.}
\label{xbeatfit} 
\end{figure}

In Figure \ref{xbeatfit} we show model hard X-ray light curves
over-laid on top of the hard X-ray data. The model light curves have
been normalised to the data. We stress that the model light curves are
a prediction of the hard X-ray flux and have not been fitted to the
data. Good fits are, therefore, not expected but where there are major
differences between data and model we can arrive at some tentative
conclusions.

Firstly, the general agreement between the model and the data is
rather good. In particular, the maximum peak in the model X-ray light
curves agrees very well with the shift in the observed peak from one
beat phase to the next. This suggests that brightest X-ray and
polarised emission regions are indeed closely coincident. A closer
inspection of beat phase $\psi =0.56$ shows that the model has
over-estimated the amount of X-ray flux at spin phases
$\phi$=0.1--0.4. This can either imply that the X-rays are being
absorbed, perhaps by the accretion stream, or the X-ray emitting
region is somewhat less extended than the cyclotron emission
region. Examining Figure \ref{globes1}, it is not clear that the
orientation of the stream is such that the X-rays from the accretion
region would be heavily absorbed at spin phases $\phi$=0.1--0.4. On
the other hand, the actual extension of the cyclotron emission region
is difficult to determine accurately at the moment due to the low time
resolution of the polarimetric data. Similarly for beat phases $\psi
=0.02$ and $\psi =0.17$ there is excess model X-ray flux at certain
spin phases.

We now go on to examine the X-ray spectrum of RX J2115--5840. A mean
background subtracted PCA X-ray spectrum covering 21--27 July 1998 was
extracted over the 2-15keV energy range and binned to improve the
signal to noise. The spectrum was fitted with a range of models
consisting of variations of an absorbed hot thermal plasma (eg a cold
absorber plus thermal bremsstrahlung, a cold absorber and partially
ionised absorber plus a hot thermal plasma, cf Cropper, Ramsay \& Wu
1998). A good fit could not be achieved -- the best fit gave a
\rchi=2.12 (31 dof). The largest source of error between the models
and the data were at energies less than 4keV. At these energies,
absorption effects at the base of the post-shock flow, where it merges
with the white dwarf, become important and are difficult to
model. This is consistent with the results discussed above which
suggest that at certain beat and spin phases, the effects of
absorption is evident.

To make some progress in extracting information from our X-ray
spectrum, we excluded data softer than 4keV. At energies harder than
this the absorbing column can be modelled using a simple cold absorber
rather than a more complex one such as a partial covering model. While
the data could be adequately modelled using an absorbed optically thin
plasma, (such as the {\tt MEKAL} model), we chose to use a
multi-temperature model (Wu, Chanmugam \& Shaviv 1994) as implemented
by Cropper, Ramsay \& Wu (1998) and improved to take into account the
effects of gravity over the height of the post-shock flow (Cropper et
al 1999). This model allows us to determine parameters such as mass of
the white dwarf and the local mass accretion rate.  Our model also
takes into account the fact that some fraction of the accretion
luminosity is radiated in optical/IR wavelengths through cyclotron
radiation. The relative proportion of the cyclotron component
increases with increasing magnetic field strength.

\begin{figure}
\begin{center}
\setlength{\unitlength}{1cm}
\begin{picture}(8,6)
\put(-7.5,-0.5){\includegraphics{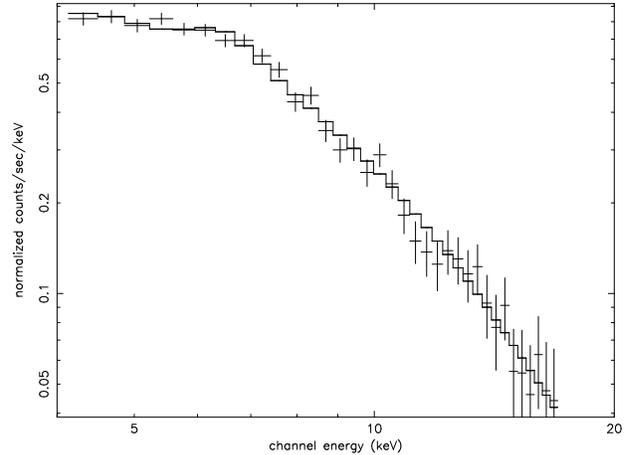}}
\end{picture}
\end{center}
\caption{The fit to the mean background subtracted {\sl RXTE} PCA 
X-ray spectrum. The model is an absorbed multi-temperature
model -- see text for details.} 
\label{xrayplot} 
\end{figure}

Using the above multi-temperature model and a cold absorber a good fit
to the data was achieved (\rchi=1.16: 23 degrees of freedom). In this
model, the ratio of cyclotron to thermal bremsstrahlung cooling,
$\epsilon_{o}$, was fixed at 1.0 at the shock: this gave an inferred
magnetic field strength at the accretion region of $B\sim$15 MG.
Since the accretion regions are offset from their respective magnetic
poles, we do not expect the field at the accretion regions to match
the dipole field strength. However the resulting fit is not very
sensitive to this parameter and almost unchanged solutions result from
field strengths between 10--20 MG. Our chosen value matches that used
for our polarimetry model fits. The viewing angle was fixed at
30$^{\circ}$ although this parameter has a very small effect on the
resulting white dwarf mass (varying this parameter between
0--90$^{\circ}$ changes the mass by 0.01 $M_{\odot}$). Figure
\ref{xrayplot} shows the fit to the data using this model while Table
\ref{xrayfits} shows the best fit parameters.

\begin{table}
\begin{center}
\begin{tabular}{lr}
\hline
$N_{H}$ & $7.3\times10^{20}$ cm$^{-2}$\\
$\dot{m}$& 0.99 g s$^{-1}$ cm$^{-2}$\\
Z & 0.53 solar\\
$M_{1}$& 0.79 (0.68--0.91) $M_{\odot}$\\
\rchi &1.16 (23 dof)\\
Observed flux & $8.5\times10^{-12}$ erg s$^{-1}$ cm$^{-2}$\\
Unabsorbed flux & $9.9\times10^{-12}$ erg s$^{-1}$ cm$^{-2}$\\
\hline
\end{tabular}
\end{center}
\caption{The best fits to the mean background subtracted PCA 
spectrum (4--15keV). The model is an absorbed multi-temperature
thermal bremsstrahlung model. $Z$ is the metal abundance relative
to solar, $M_{1}$, is the mass of the white dwarf and the values
in brackets refer to the 90 per cent confidence range. The two flux 
measurements are the implied flux over the 4--15keV energy range.}
\label{xrayfits}
\end{table}

Ramsay et al (1999) suggested that their polarimetry data implied the
accretion regions had different magnetic field strengths, indicating
evidence for a dipole offset or a magnetic field geometry more complex
than a simple dipole.  This is consistent with the observation of
Schwope et al (1997) who found that at one epoch RX J2115 was bright
in the EUV and at another epoch much stronger in hard X-rays. If there
was a large dipole offset then it is expected that there would be a
significant difference in the hard X-ray spectra of the two accreting
poles.

We extracted the X-ray data originating from each pole by assuming
that at epochs when only positive circular was detected accretion was
occurring on to only one pole. When only negative polarisation was
detected accretion was occurring onto the other pole. These two
spectra were fitted using the same best fit model to the mean spectrum
with only the normalisation being allowed to vary. Good fits
(\rchi$<$1.05) were obtained to both spectra indicating that there is
no evidence (from these data) for a significant difference in the hard
X-ray spectra from each accreting pole. The data from each night were
also fitted with the best fit model and equally good fits were
found. We conclude that these X-ray data are consistent with a simple
dipole magnetic field as used for the polarisation modelling.

\section{The moving accretion regions in RX J2115--58}
\label{shift}

Ramsay et al (1999) presented polarisation data covering a large
fraction of the spin-orbit beat interval. Using these data they
proposed a very preliminary model in which the accretion flow was
directed onto one or other footprint of the same magnetic field line
at all beat phases. The main accretion regions on opposite
hemispheres were fixed to within $\sim70^{\circ}$ in magnetic
longitude. With the detailed modelling of our new data presented here,
it is possible to test the preliminary model of Ramsay et al (1999).

We first concentrate on the accretion pattern in the lower hemisphere,
in which accretion occurs at all beat phases where we have been able
to Stokes Image the data. Figure \ref{pol_model2} shows that at beat
phase $\psi\sim$0.55 an extended accretion region is seen in this
hemisphere.  As we increase in beat phase, we find that the brightest
(strictly speaking the most dense) part of the accretion region
advances in positive longitude: between beat phase $\psi$=0.55 and
0.85, we find it has increased in spin longitude by $\sim50^{\circ}$
from $\sim220-270^{\circ}$. This shift is also seen in the advancement
in spin phase of the peak of the positive circular polarisation in
Fig. \ref{pol_model} (also seen in Ramsay et al 1999), and is
consistent with our X-ray data. During this interval the region also
splits into two distinct components separated by $\sim180^{\circ}$ in
longitude -- this has clearly occurred by beat phase
$\psi$=0.00. Slightly later, at beat phase $\psi=0.15$, for a short
phase interval, the originally most prominent region in the lower
hemisphere is almost absent, and accretion occurs onto the upper
hemisphere. This new region is located near the longitude of the lower
magnetic pole. At this beat phase, the two accretion regions are
therefore approximately $180^{\circ}$ in longitude from their
respective magnetic poles. Alternatively, they are at the expected
longitudes, but at the opposite latitudes to that expected. We are
unable to account for this behaviour at this beat phase. However, it
is clear that accretion never occurs near the upper pole, whatever the
azimuth of the magnetic poles. This probably indicates the presence of
a non-dipole field with the field strength at the upper pole
significantly higher, in agreement with the suggestion of Schwope et
al (1997) and despite the lack of evidence from our X-ray data.

Ramsay et al (1999) suggested that at beat phase $\psi=0.00$ it is
possible that the stream divides close to the white dwarf so that one
stream accretes most directly and the other follows a path around the
white dwarf. Alternatively, and in our view more likely, the stream
separates much closer to the secondary star and follows a path some
distance in azimuth around the white dwarf before accreting. As we
move further in beat phase (to $\psi$=0.15), we find the accretion
region which was located at $270^{\circ}$ in longitude (Figure
\ref{pol_model2}) has become much reduced and the accretion flow onto
the lower hemisphere is directed onto the accretion spot at
$90^{\circ}$ in longitude. It is also at this point in the beat cycle
that accretion is directed onto the upper hemisphere. By the time of
our next accretion map ($\psi$=0.55), we find that the first accretion
footprint is again the most favourable and this is where the bulk of
the accretion flow is directed.

We return finally to the cycle-to-cycle variability of the circular
polarisation evident in Figure \ref{cpol1}. This appears to be most
marked at the beat phases at which the circular polarisation is
predominantly negative, ie those beat phases around $\psi=0.15$ when
the accretion occurs in the upper hemisphere. The variability is
therefore most likely caused by changes in the balance of the
accretion flow to upper and lower hemispheres via the two streams,
resulting in a net polarisation of approximately zero.

\section{Acknowledgments}

We would to thank the Director of SAAO, Dr R Stobie, for the generous
allocation of observing time and we are grateful to Dr D O'Donoghue
for the use of his period analysis software.

\end{document}